\documentclass[sigconf]{acmart} 
\AtBeginDocument{%
  }

\setcopyright{acmlicensed}
\copyrightyear{2026}
\acmYear{2026}
\acmDOI{}
\acmConference[AI-SQE '26]{The 1st International Workshop on AI for Software Quality Evaluation: Judgment, Metrics, Benchmarks, and Beyond}{April 14,
  2026}{Rio de Janeiro, Brazil}
\acmISBN{}

\usepackage{url}
\usepackage{subcaption}
\usepackage{graphicx}
\usepackage{listings}
\usepackage{stfloats}
\usepackage{eso-pic}
\lstdefinestyle{mystyle}{
    basicstyle=\fontsize{7}{7}\selectfont\rmfamily,
    columns=fullflexible, 
    frame=single,          
    rulecolor=\color{black},
    breaklines=true,
    aboveskip=0pt,
    belowskip=0pt,
    breakindent=0pt,
    moredelim=[is][\textcolor{blue}]{*blue*}{*blue*},
    moredelim=[is][\textcolor{red}]{*red*}{*red*},
    moredelim=**[is][\bfseries]{*bold*}{*bold*}, 
    literate={```}{\textasciigrave\textasciigrave\textasciigrave}1,
}
\definecolor{lightergray}{RGB}{240,240,240}

\newcommand{\mybox}[1]{
\vspace{0.5em}
\noindent
\fcolorbox{black}{lightergray}{\parbox{.975\columnwidth}{#1}}
\vspace{0.5em}
}

\begin{document}

\AddToShipoutPictureBG*{%
  \AtPageUpperLeft{%
    \put(40,-555){
      \rotatebox{90}{%
        \makebox[0pt][l]{%
          \large\sffamily\color{gray}%
          Accepted at ICSE 2026: AI for Software Quality Evaluation (AI-SQE) Workshop%
        }%
      }%
    }%
  }%
}

\title[On the Quality of AI-Generated Source Code Comments]{On the Quality of AI-Generated Source Code Comments:\\ A Comprehensive Evaluation}

\author{Ian Guelman}
\email{ianguelman@dcc.ufmg.br}
\affiliation{%
  \institution{Department of Computer Science, Federal University of Minas Gerais (UFMG)}
  \city{Belo Horizonte}
  \state{Minas Gerais}
  \country{Brazil}
}

\author{Arthur Gregório Leal}
\email{arthurgregorioleal@gmail.com}
\affiliation{%
  \institution{Department of Software Engineering, Pontifical Catholic University of Minas Gerais (PUC Minas)}
  \city{Belo Horizonte}
  \state{Minas Gerais}
  \country{Brazil}
}

\author{Laerte Xavier}
\email{laertexavier@pucminas.br}
\affiliation{%
  \institution{Department of Software Engineering, Pontifical Catholic University of Minas Gerais (PUC Minas)}
  \city{Belo Horizonte}
  \state{Minas Gerais}
  \country{Brazil}
}

\author{Marco Tulio Valente}
\email{mtov@dcc.ufmg.br}
\affiliation{%
  \institution{Department of Computer Science, Federal University of Minas Gerais (UFMG)}
  \city{Belo Horizonte}
  \state{Minas Gerais}
  \country{Brazil}
}


\begin{abstract}
This paper investigates the quality of source code comments automatically generated by Large Language Models (LLMs). While AI-based comment generation has emerged as a promising solution to reduce developers’ documentation effort, prior studies have been limited by small datasets or by relying solely on traditional Information Retrieval (IR) metrics, which are insufficient to capture documentation quality. To address these limitations, we conducted a large-scale empirical study on 142 classes and 273 methods created after the training cut-off of the evaluated models. For each code element, we generated Javadoc comments using three LLMs (GPT-3.5 Turbo, GPT-4o, and DeepSeek-V3). A qualitative assessment of the comments---performed independently by two experts---showed that 58.8\% were equivalent to, and 27.7\% superior to, the original comments. A quantitative analysis using BLEU, ROUGE-L, and METEOR confirmed that IR-based metrics do not reliably reflect human evaluations, revealing the need for new documentation-specific metrics. Finally, correlation analyses indicated slightly positive relationships between code properties (size, complexity, coupling) and comment quality, confirming that LLMs benefit from richer contextual information.
\end{abstract}



\received{20 October 2025}
\received[Accepted]{23 Nov 2025}

\maketitle

\section{Introduction}

Documentation plays a key role in software development. In particular, source code comments remain the most common type of software documentation~\cite{GPT3Documentation}, used to enhance readability, facilitate maintenance, and reduce development effort~\cite{commentsGenerationSurvey, LLMDocumentationAnalysis, canDevsPrompt, softengbook}. However, because writing comments is perceived as time-consuming, tedious, and an interruption to the coding flow, many developers neglect this practice. To address this problem, AI-driven automated source comment generation---particularly through Large Language Models (LLMs)---offers a compelling solution that helps practitioners bridge the gap between efficient coding and effective documentation~\cite{GPT3Documentation, practitionersExpectations}.

However, we still lack studies that provide a comprehensive evaluation of the quality of AI-generated source code comments. Existing research often relies on small samples, such as the study by \citet{LLMDocumentationAnalysis}, in which the authors evaluated only 14 comments. Other studies depend solely on classical Information Retrieval (IR) metrics, which may fail to capture the nuances of code documentation. For example, \citet{GPT3Documentation} assessed the quality of LLM-generated comments using a single IR-based metric (BLEU). Finally, some studies focus on code summarization, which aims to help developers understand a defective or extensible block of code. Typically, such studies---like the one by \citet{llmEra}---evaluate the use of LLMs to generate a single sentence explaining a snippet of code to human maintainers.

To provide a more comprehensive evaluation of the quality of source comments generated by LLMs, this paper presents a study with the following characteristics:

\begin{enumerate}
\item The study relies on a dataset comprising 142 classes and 273 methods, along with their associated Javadocs.
All selected classes and methods are located in files created after the training cut-off date of the LLMs used in the study, ensuring that the models could not reuse previously known comments.
For each selected class and method, we generated Javadoc comments using three LLMs (OpenAI GPT-3.5 Turbo, OpenAI GPT-4o, and DeepSeek-V3). The dataset is publicly available at \url{https://doi.org/10.5281/zenodo.17343478}.

\item The study includes a qualitative evaluation of the comments generated by one of the selected LLMs (GPT-3.5 Turbo). Two authors independently assessed the quality of the generated comments using a four-point Likert scale. In cases of disagreement, a meeting was held to reach a consensus score. The results indicate that most LLM-generated comments are of high quality, according to the expert evaluators: 58.8\% were rated as equivalent to the original comments, and 27.7\% were rated as superior.

\item The study includes a quantitative evaluation of the comments generated by the selected LLMs using three IR-based metrics (BLEU, ROUGE-L, and METEOR). As one of our key contributions, we found that these metrics are not a substitute for human judgment, as they fail to capture the nuanced differences inherent in source code comments. Consequently, this finding highlights the need for new metrics specifically designed to assess the quality of code documentation generated by LLMs.

\item Finally, the sudy includes a correlation analysis involving two groups of metrics: human evaluation scores and source code metrics that measure properties such as size, complexity, cohesion, and coupling. We found slight positive correlations between size, complexity, and number of dependencies, and the quality of the generated comments (Spearman $\leq$ 0.30). In other words, this findings confirms that LLMs require context---in the form of lines of code or dependencies---to generate high-quality comments.
\end{enumerate}

The remainder of this paper is organized as follows. Section~\ref{sec:studyDesign} describes the study design. Sections~\ref{sec:qualitativeAssessment} and~\ref{sec:quantitativeAssessment} present the qualitative and quantitative evaluations, respectively. Section~\ref{sec:correlating} correlates the results of these two evaluations, and Section~\ref{sec:codeAssessment} relates the results to CK metrics. Section~\ref{sec:threatsToValidity} discusses threats to validity and 
Section~\ref{sec:relatedWork} discusses related work.  Section~\ref{sec:conclusionAndFutureWork} concludes.

\section{Study Design}
\label{sec:studyDesign}
Our study employs a comprehensive approach to systematically evaluate the performance of LLMs in generating Javadoc comments. To achieve this, we utilized a combination of established natural language processing metrics, human evaluation, and a collection of well-known source code metrics to perform both quantitative and qualitative assessments. 


\subsection{Dataset}
We selected three widely adopted Java-based GitHub repositories: {\sc spring-projects/spring-boot}, {\sc spring-projects/spring-frame\-work}, and {\sc google/guava}. These projects were chosen for their relevance, sustained developer activity, and strong community support---each with over 50,000 stars. Collectively, they represent core aspects of Java development, making them well-suited for analyzing Java source code documentation.

To mitigate implicit data leakage from pre-training, we followed the guidelines by Sallou, Durieux, and Panichella~\cite{guidelines}. Since the repositories are open source and may have been part of the LLM’s training data, we selected only files created after the model’s training cut-off date. From these, we extracted Javadoc comments and their corresponding code.



Following Khan and Uddin~\cite{GPT3Documentation}, we filtered out comments that: (i) were not in English; (ii) contained special tokens or external links (e.g., \texttt{<img>} and \texttt{https://}); or (iii) included author metadata (i.e., \texttt{@author} tag). This ensures that AI-driven comment generation relies solely on code-derived context and that it preserves privacy.

\subsection{Ground Truth}
\label{subsec:groundTruth}
Since human evaluation is both time-consuming and costly~\cite{commentsGenerationSurvey}, we selected a representative sample of our dataset to serve as our ground truth. 
To achieve a statistically significant sample, based on a 95\% confidence level and a 5\% margin of error, we randomly selected 142 class-level and 273 method-level Javadocs. As part of an initial data curation step, these original comments were carefully evaluated by two of the authors to assess their quality and consistency with the code they document. In other words, our goal was to prevent low-quality original comments from being included in our ground truth. Thus, if either author considered a comment inaccurate in describing its corresponding code element, it was discarded and replaced with another randomly selected comment. Only 6\% of the Javadocs required replacement, demonstrating that the proposed dataset has a high level of quality.

\subsection{Javadoc Generation Using LLMs}
\label{subsec:javadocGeneration}

For each method and class in our ground truth, we extracted the corresponding code and used only it to  prompt the LLMs to generate a Javadoc. In every case, we provided the LLM with only the code contained within the target entity (i.e., the method or class body), omitting all external context such as imports and surrounding code elements. The process of prompting the LLM was done programmatically using the available APIs. The prompt employed is shown in Figure~\ref{fig:prompt}. We analyze the LLM-generated Javadocs in this study.

\begin{figure}[htb]
\centering
\begin{lstlisting}
Context: Suppose you are a Java developer who needs to document some 
*blue*[`methods' | `classes']*blue* using source code comments following the Javadoc format.
For example, for the following *blue*[`method' | `class']*blue*:
```java
*blue*[code example]*blue*
```
You should generate the following Javadoc comment:
*blue*[javadoc example]*blue*

Task: Generate a single comment (in Javadoc format) for the following 
*blue*[`method' | `class']*blue*
```java
*blue*[extracted code]*blue*
```
In your answer, only include the suggested comment.
\end{lstlisting}
\caption{Prompt structure}
\label{fig:prompt}
\end{figure}

\subsection{Qualitative Analysis}

To assess the effectiveness of the LLMs in generating Javadocs, we first conducted a human evaluation, comparing the LLM-generated Javadocs with their ground-truth counterpart. Following the approach of Tran et al.~\cite{bleuLikert}, we introduced the Human Assessment Score (HA-Score), a four-point scale defined as follows: 

\begin{itemize}
    \item A score of 4 indicates that the generated comment is fully equivalent to the original, conveying the same information without requiring modifications;
    \item A score of 3 means that the generated comment is mostly correct but requires minor modifications to achieve equivalence with the original;
    \item A score of 2 denotes that the generated comment is somewhat usable but requires significant revisions to align with the original;
    \item A score of 1 implies that the generated comment is entirely incorrect and cannot serve as a replacement for the original.
\end{itemize}

Two of the authors independently evaluated each generated Javadoc, assigning an HA-Score as defined above (142 class-level and 273 method-level Javadocs, totaling 415 Javadocs × 2 authors = 830 evaluations).
To assign the scores, each evaluator compared the original and the LLM-generated Javadoc and also examined the corresponding source code element (function or class), if needed.
This analysis of the code was particularly important to justify a score of 1 (when the generated comment was completely incorrect) and to confirm cases where the LLM-generated comment was indeed superior to the original.
The latter cases were assigned a score of 4+, as detailed in Section~\ref{sec:qualitativeAssessment}.

Both evaluators are experienced software engineers with 6 and 7 years of experience, having authored numerous documentations for various projects. To minimize biases, the authors also discussed their individual evaluations and resolved disagreements to reach a consensus on a final score for each pair, which is the one referenced throughout this paper. The evaluation yielded a Cohen’s Kappa value of 0.54, indicating moderate agreement~\cite{kappaRange}. However, only 3\% of the disagreements differed by more than one point, indicating a minimal impact of disagreement on the overall assessment.

\subsection{Quantitative Analysis}

For the quantitative analysis, we measured textual similarity using BLEU~\cite{bleu}, ROUGE-L~\cite{rouge}, and METEOR~\cite{meteor}. BLEU is a precision-oriented metric that measures n-gram overlap between two texts. While it is widely used for evaluating machine-generated text~\cite{codeSummarization, bleuGuilty}, it has well-known limitations and may not always provide accurate assessments~\cite{bleuGuilty, tangledBleu, reassessingMetrics}. To enhance its reliability, we applied a smoothing function following Khan and  Uddin~\cite{GPT3Documentation}, making BLEU more sensitive to partial matches. 

Additionally, to complement BLEU, we incorporated ROUGE-L---a recall-oriented metric based on the longest common subsequence. Unlike BLEU, ROUGE-L does not rely on exact matches, making it suited for capturing sentence-level structural similarity. We also employed METEOR---a balanced metric that combines recall with precision by incorporating stemming, synonym matching, and paraphrase recognition. This characteristic makes METEOR more robust in evaluating textual similarity.

\subsection{Source Code Metrics Analysis}

To investigate which source code properties influence LLMs' performance in Javadoc generation, we extracted source code metrics using the CK Tool~\cite{aniche-ck}, which computes various metrics, including the Chidamber and Kemerer (CK) suite~\cite{ckMetrics}. We selected all CK metrics except Depth of Inheritance Tree (DIT) and Number of Children (NOC), as these refer to class hierarchy rather than the code directly provided to the LLM. Additionally, we incorporated Lines of Code (LOC) and Unique Words Quantity, both relevant to how LLMs process textual input. Thus, the final metric set includes Weighted Methods per Class (WMC), Coupling Between Objects (CBO), Response for a Class (RFC), Lack of Cohesion of Methods (LCOM), LOC, and Unique Words Quantity.

\subsection{Metadata}
\label{subsec:metadata}

To enhance reproducibility, we follow LLM research guidelines~\cite{guidelines} and report detailed prompting metadata.\\[-0.25cm]

\noindent \textbf{Model.} For the qualitative analysis, we used OpenAI GPT-3.5 Turbo (gpt-3.5-turbo-0125), as it offered the best cost-benefit ratio at the time of the study. For the quantitative analysis, we incorporated two more recent models: OpenAI GPT-4o (gpt-4o-2024-08-06) and DeepSeek-V3 (deepseek-chat), to provide a more comprehensive and up-to-date evaluation.\\[-0.25cm]


\noindent \textbf{Prompt.} As already presented in Section~\ref{subsec:javadocGeneration}, we adopted a one-shot learning approach, providing a single example with task-specific instructions to guide the model in generating Javadocs from a code snippet. This strategy was selected after preliminary experiments comparing zero-shot, one-shot, and few-shot prompts. One-shot prompting consistently produced better results, while adding more examples (few-shot) showed no significant gains. 

\section{Qualitative Results}
\label{sec:qualitativeAssessment}

For the qualitative assessment, we evaluated the quality of the Javadoc generated by GPT-3.5 Turbo. 
We selected this model because it offered the best cost-effectiveness at the time the study was conducted. This decision also allows us to position our results as a lower bound. That is, newer and more advanced models are expected to perform even better, producing higher-quality Javadocs.

\begin{figure}[!htb]
    \centering
    \includegraphics[width=\linewidth]{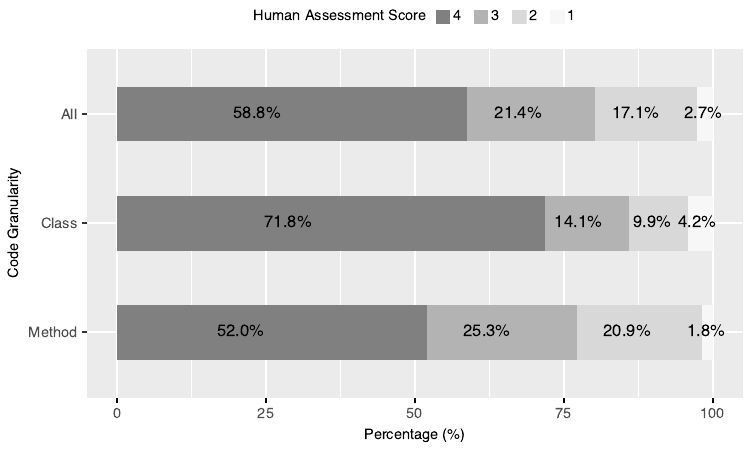}
    \caption{Human assessment scores}
    \label{fig:hasDistribution}
\end{figure}

The evaluation involved assigning a HA-Score to each GPT-generated Javadoc.
As can be seen in Figure~\ref{fig:hasDistribution}, the majority of the generated comments (58.8\%) received a HA-Score of 4, indicating equivalence with the original Javadoc. This proportion increased to 71.8\% when analyzing classes specifically. Interestingly, some generated comments not only matched the original Javadoc in quality---receiving a score of 4---but also enhanced the information they conveyed. To better capture this fact, we introduce a new score, denoted as 4+, which includes comments that surpass the original Javadoc in content. These accounted for nearly half (47.1\%) of the Javadocs rated at 4 and comprised 27.7\% of the entire sample.\vspace*{2mm}

\mybox{\textbf{Finding \#1:} The comments generated by LLMs are of high quality, according to the experts responsible for the qualitative assessment. The majority of AI-generated comments (58.8\%) were rated as equivalent to the original ones. Additionally, a significant portion (27.7\%) were judged to be of higher quality than the originals.}

\subsection{Human Assessment Scoring Examples}

                To enhance clarity and understanding of our results, we present an example of the generated Javadoc for each score, alongside the original Javadoc that documented the respective code snippet from which the generated Javadoc was derived. We then discuss the reasons behind each classification.

                Both Javadocs in Figure~\ref{fig:hasExample4+} describe an object that captures other objects to be closed when a ClosingFuture pipeline is completed. However, the generated comment is more comprehensive, offering additional details about the class, which enhances understanding from the documentation alone. This additional, useful information merits the distinguished Score of 4+. Javadocs achieving this score consistently provide more valuable insights than the originals, enriching the reader's comprehension of the code.
            
                \begin{figure}[!htbp]
                    \centering
                    \begin{lstlisting}
*bold*Original*bold*
/**
* An object that can capture objects to be closed later, when a {@link ClosingFuture} pipeline is
* done.
*/

*bold*LLM-Generated*bold*
/**
* A class that captures objects to be closed when a ClosingFuture pipeline is done.
*
* <p>Objects can be any AutoCloseable for the -jre flavor of Guava or Closeable for the -android flavor.
* It is important to ensure compatibility with older SDK versions by passing objects that implement the necessary
* interfaces at the oldest supported version.
*
* <p>The method is binary-compatible between flavors due to the parameter type erasure being Object.
*
* @see CloseableList
*/
                    \end{lstlisting}
                    \caption{Human Assessment Score---Example Score 4+}
                    \label{fig:hasExample4+}
                \end{figure}

                The Javadocs in Figure~\ref{fig:hasExample4} are nearly identical, differing only in subtle phrasing variations, thus justifying a score of 4 for the generated comment. This score typically applies to Javadocs that are almost exact replicas of the originals, with only slight linguistic differences.
                
                \begin{figure}[!htbp]
                    \centering
                    \begin{lstlisting}
*bold*Original*bold*
/**
* Register a {@link DeferredResultProcessingInterceptor} under the given key.
* @param key the key
* @param interceptor the interceptor to register
*/

*bold*LLM-Generated*bold*
/**
* Register a {@link DeferredResultProcessingInterceptor} with a given key.
* 
* @param key the key to associate with the interceptor
* @param interceptor the interceptor to be registered
*/
                    \end{lstlisting}
                    \caption{Human Assessment Score---Example Score 4}
                    \label{fig:hasExample4}
                \end{figure}

                Both Javadocs in Figure~\ref{fig:hasExample3} describe the method's functionality: reversing an array's elements. However, the generated comment is concise, whereas the original elaborates on an equivalent method and its efficiency. We considered that the omissions are minor and do not hinder understanding of the method's purpose. Therefore, it receives a score 3.
            
                \begin{figure}[H]
                    \centering
                    \begin{lstlisting}
*bold*Original*bold*
/**
* Reverses the elements of {@code array}. This is equivalent to {@code
* Collections.reverse(Longs.asList(array))}, but is likely to be more efficient.
*
* @since 23.1
*/

*bold*LLM-Generated*bold*
/**
* Reverse the elements in the given array.
* @param array the array to be reversed
*/
                    \end{lstlisting}
                    \caption{Human Assessment Score---Example Score 3}
                    \label{fig:hasExample3}
                \end{figure}

                 Both Javadocs in Figure~\ref{fig:hasExample2} discuss a waiting time before shutdown. The original comment, however, provides additional details (e.g., behavior when tasks are submitted, and default wait periods). The generated version, which lacks these important details, requires major additions to reach equivalence, thus warranting a score of 2.
                
                \begin{figure}[!htbp]
                    \centering
                    \begin{lstlisting}
*bold*Original*bold*
/**
* Configure the amount of time we'll wait before shutting down resources.
* If a task is submitted during the {@code shutdownQuietPeriod}, it is guaranteed
* to be accepted and the {@code shutdownQuietPeriod} will start over.
* <p>By default, this is set to
* {@link LoopResources#DEFAULT_SHUTDOWN_QUIET_PERIOD} which is 2 seconds but
* can also be overridden with the system property
* {@link reactor.netty5.ReactorNetty#SHUTDOWN_QUIET_PERIOD
* ReactorNetty.SHUTDOWN_QUIET_PERIOD}.
* @since 5.2.4
* @see #setShutdownTimeout(Duration)
*/

*bold*LLM-Generated*bold*
/**
* Set the duration of the quiet period before shutdown.
* @param shutdownQuietPeriod the duration of the quiet period
*/
                    \end{lstlisting}
                    \caption{Human Assessment Score---Example Score 2}
                    \label{fig:hasExample2}
                \end{figure}

               The original Javadoc in Figure~\ref{fig:hasExample1} describes a method that only throws an exception. The LLM similarly notes the exception, but incorrectly claims that all map values will be replaced. This discrepancy justifies a score 1, indicating a fundamental misunderstanding of the method, a common trait among other Javadocs ranked similarly.

                \begin{figure}[H]
                    \centering
                    \begin{lstlisting}
*bold*Original*bold*
/**
* Guaranteed to throw an exception and leave the map unmodified.
*
* @throws UnsupportedOperationException always
* @deprecated Unsupported operation.
*/

*bold*LLM-Generated*bold*
/**
* Replaces all key-value pairs in the map by applying the specified function, always throwing UnsupportedOperationException.
* @param function the function to apply (not used as the method always throws an exception)
* @deprecated This method is deprecated and should not be called
* @DoNotCall(""Always throws UnsupportedOperationException"")
* @throws UnsupportedOperationException always thrown as this operation is not supported
*/
                    \end{lstlisting}
                    \caption{Human Assessment Score---Example Score 1}
                    \label{fig:hasExample1}
                \end{figure}

\section{Quantitative Results}
\label{sec:quantitativeAssessment}

\begin{figure*}[t]
    \centering
    \begin{tabular}{|c|c|c|c|}
        \hline
        & 
        \scriptsize \textbf{Smoothed BLEU} & 
        \scriptsize \textbf{ROUGE-L} & 
        \scriptsize \textbf{METEOR} \\ 
        \hline
        \rotatebox{90}{\scriptsize \textbf{GPT 3.5 Turbo}} & 
        \includegraphics[width=0.29\textwidth, trim=18 20 10 0, clip]{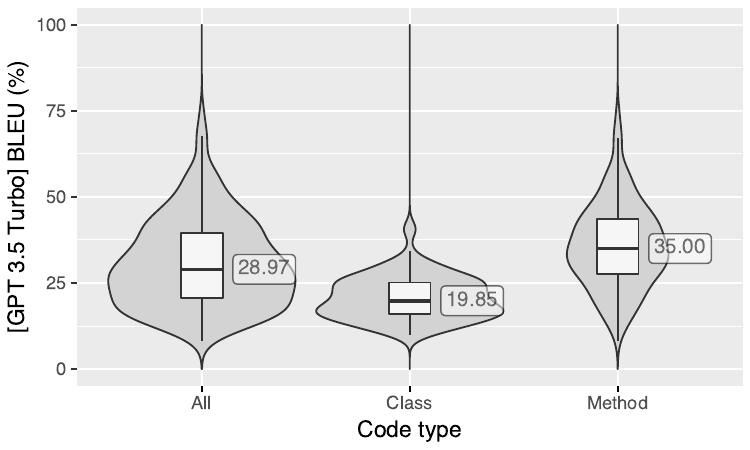} & 
        \includegraphics[width=0.29\textwidth, trim=18 20 15 0, clip]{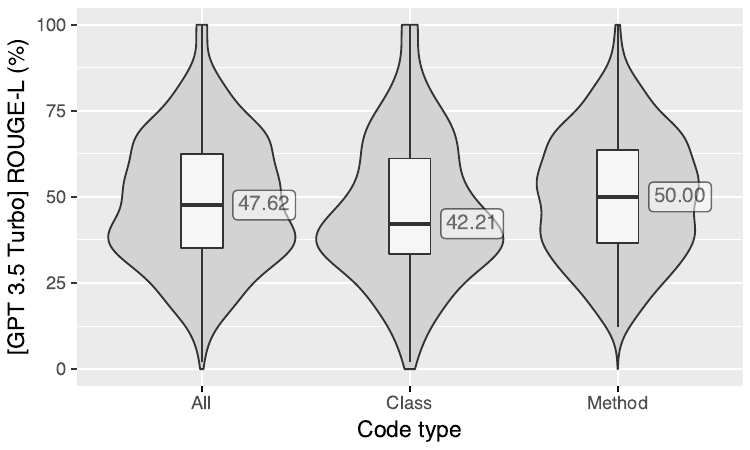} & 
        \includegraphics[width=0.29\textwidth, trim=18 20 15 0, clip]{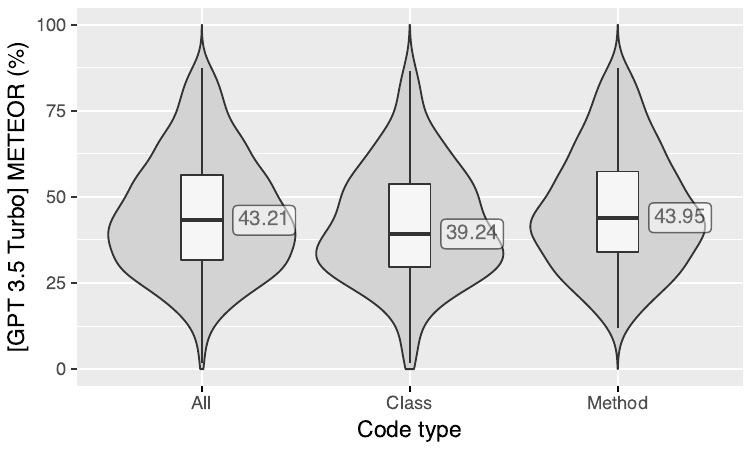} \\ 
        \hline
        \rotatebox{90}{\scriptsize \textbf{GPT 4o}} & 
        \includegraphics[width=0.29\textwidth, trim=18 20 15 0, clip]{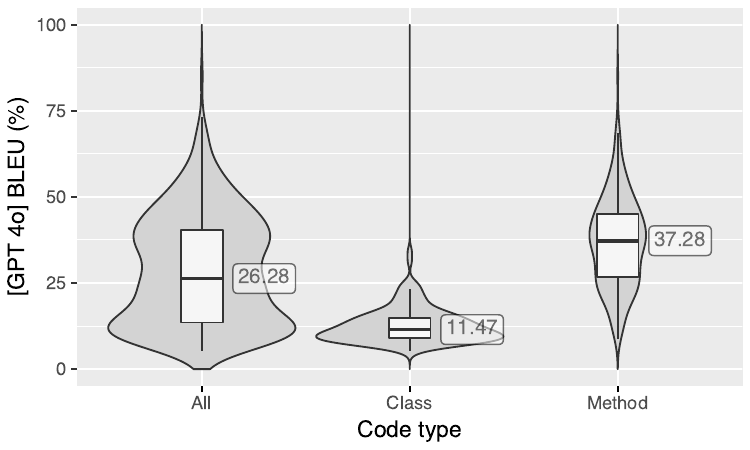} & 
        \includegraphics[width=0.29\textwidth, trim=18 20 15 0, clip]{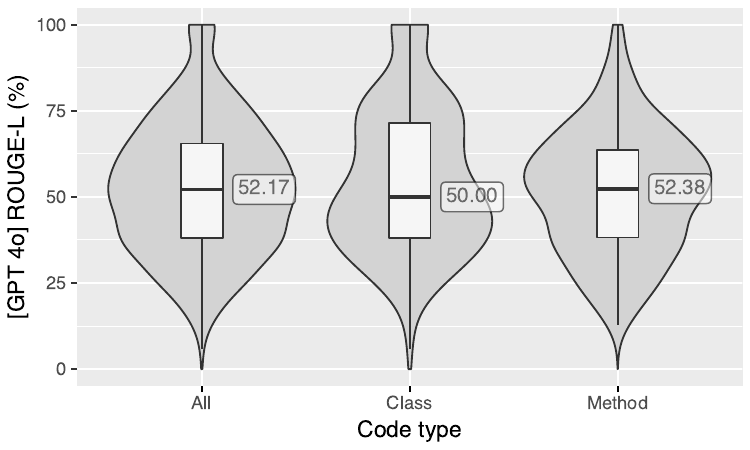} & 
        \includegraphics[width=0.29\textwidth, trim=18 20 15 0, clip]{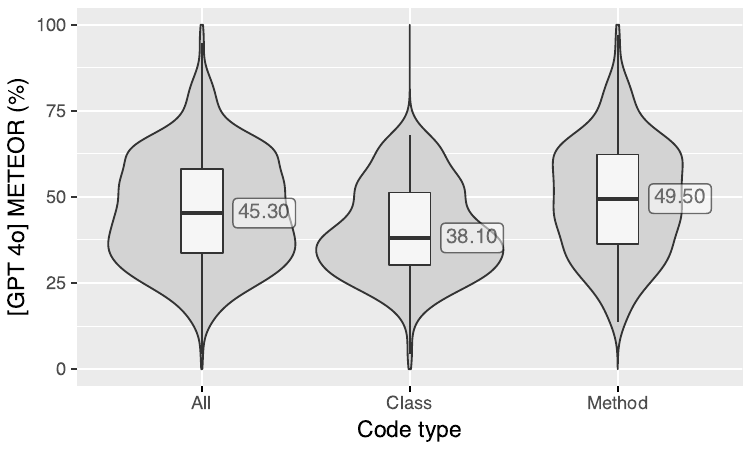} \\ 
        \hline
        \rotatebox{90}{\scriptsize \textbf{DeepSeek-V3}} & 
        \includegraphics[width=0.29\textwidth, trim=18 20 15 0, clip]{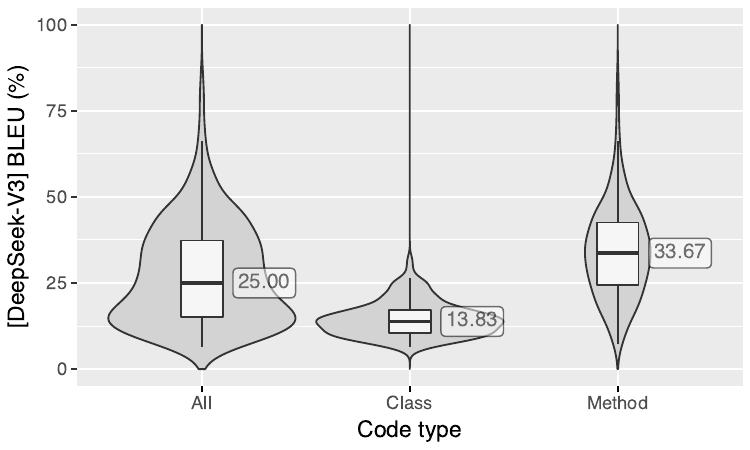} & 
        \includegraphics[width=0.29\textwidth, trim=18 20 15 0, clip]{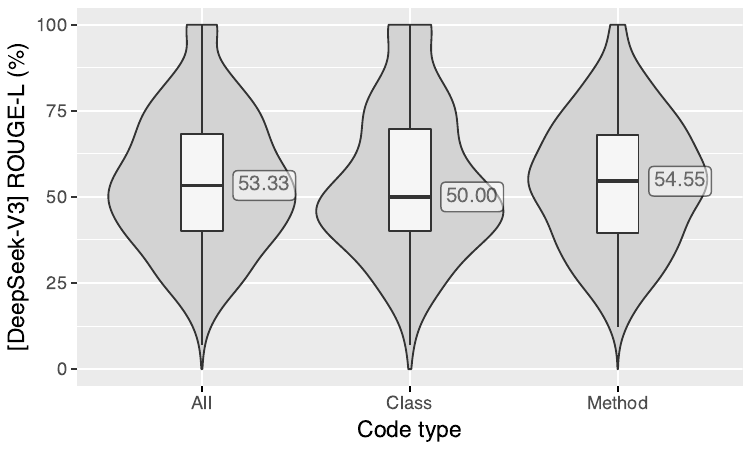} & 
        \includegraphics[width=0.29\textwidth, trim=18 20 15 0, clip]{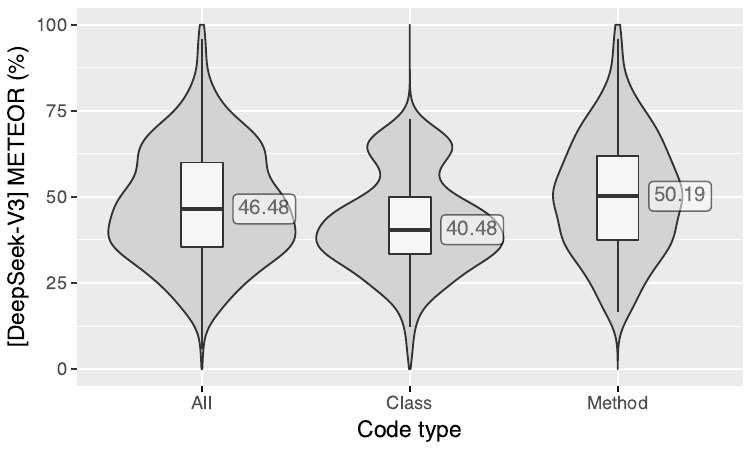} \\ 
        \hline
    \end{tabular}
    \caption{Comparison of LLMs using BLUE, ROUGE-L, and METEOR}
    \label{fig:quantitativeEvaluation}
\end{figure*}

We also performed a quantitative assessment of the Javadocs produced by the three LLMs. To evaluate their quality, we compared the generated instances against their ground-truth versions. For each pair of comments (original vs LLM-generated), we calculated BLEU, ROUGE-L, and METEOR scores to quantify their similarity and alignment.

Figure~\ref{fig:quantitativeEvaluation} presents the results obtained for the quantitative assessment. As can be seen, GPT-3.5 Turbo achieved a median BLEU score of 28.97\% (overall). GPT-4o showed a slightly lower median score (26.28\%), particularly for class-level comments (11.47\%). DeepSeek-V3 had the lowest BLEU scores, with an overall median of 25\%.
In contrast, ROUGE-L and METEOR scores revealed a different trend. DeepSeek-V3 outperformed both GPT models, with overall medians of 53.33\% and 46.48\%, respectively. GPT-4o followed closely with 52.17\% and 45.30\% overall, while GPT-3.5 Turbo lagged behind with 47.62\% and 43.21\%. 

The difference between BLEU and both the ROUGE-L and METEOR scores highlights the complementary nature of these metrics. While BLEU emphasizes exact n-gram overlaps, ROUGE-L captures longer-range sequence similarity and is more tolerant of surface-level variations in wording. In contrast, METEOR combines precision and recall while accounting for synonymy and stemming, making it recommended for evaluating semantic similarity despite lexical differences.\\[-0.25cm]

\mybox{\textbf{Finding \#2:} Newer models such as GPT-4o and DeepSeek-V3 demonstrate improved performance in generating source code comments, as evidenced by their higher ROUGE-L and METEOR scores.}

\section{Qualitative vs Quantitative Results}
\label{sec:correlating}

\begin{figure*}[!t]
    \centering
    \begin{tabular}{|c|c|c|c|}
        \hline
        & 
        \scriptsize \textbf{Smoothed BLEU} & 
        \scriptsize \textbf{ROUGE-L} & 
        \scriptsize \textbf{METEOR} \\ 
        \hline
        \rotatebox{90}{\scriptsize \textbf{GPT 3.5 Turbo}} & 
        \includegraphics[width=0.3\textwidth, trim=18 5 5 0, clip]{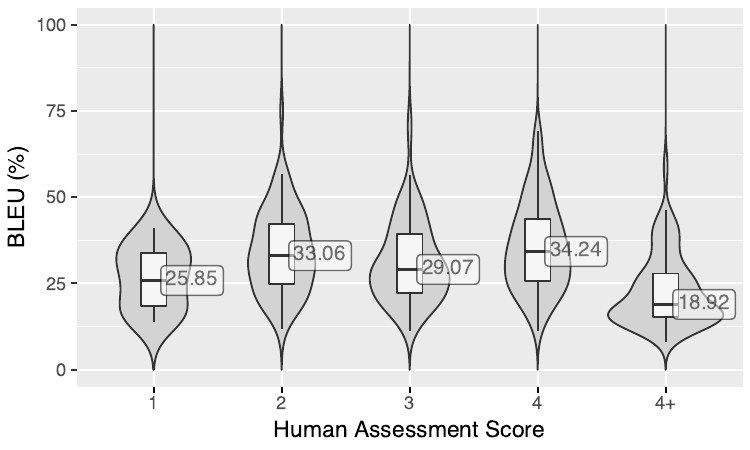} & 
        \includegraphics[width=0.3\textwidth, trim=18 5 5 0, clip]{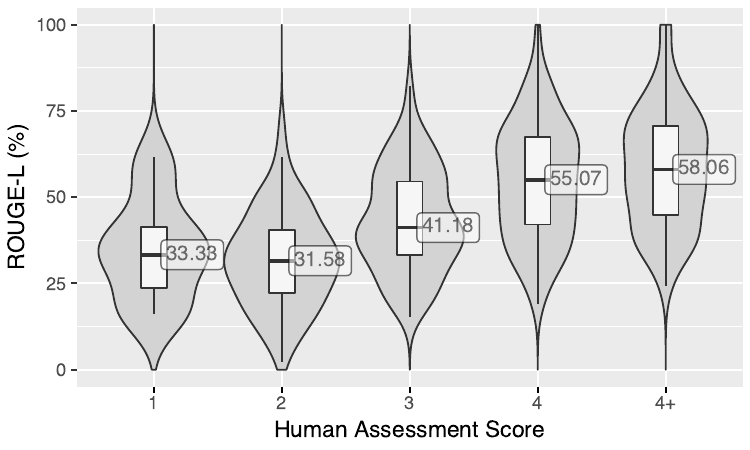} & 
        \includegraphics[width=0.3\textwidth, trim=18 5 5 0, clip]{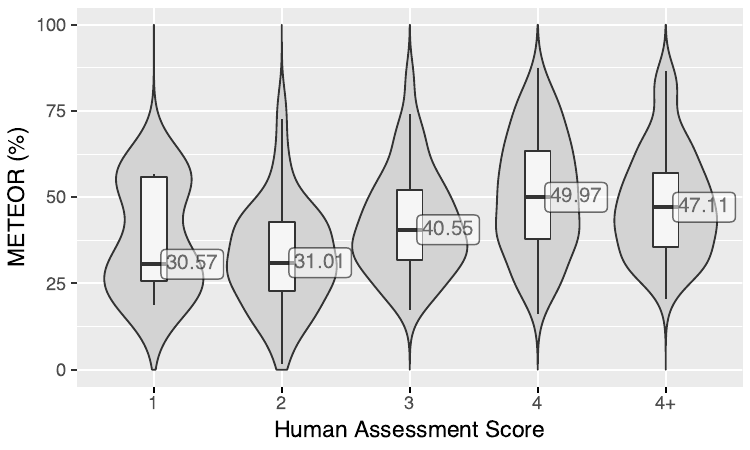} \\ 
        \hline
    \end{tabular}
    \caption{Correlation between Evaluation Metrics and HA-Score}
    \label{fig:metricsCorrelation}
\end{figure*}

When we correlate the quantitative metrics with HA-Score, we see different trends. As shown in Figure~\ref{fig:metricsCorrelation}, BLEU is the only metric that exhibits a negative correlation with the HA-Score (Spearman correlation of -0.28). In contrast, both ROUGE-L (Spearman correlation of 0.49) and METEOR (Spearman correlation of 0.30) demonstrate moderate positive correlations. These results suggest that BLEU may not adequately capture human judgment, whereas ROUGE-L and METEOR show a better alignment with human evaluation.

After analyzing the results in depth, we also concluded that conventional metrics have relevant limitations in capturing subtle differences that appear in human assessments. This is evident when we compare BLUE, ROUGE-L, and METEOR scores across the HA-Score groups. For example, none of these metrics showed statistically significant differences between HA-Score pairs ‘1’ vs. ‘2’ and ‘1’ vs. ‘3’, according to a Mann-Whithner U Test. In fact, BLEU and METEOR fail to distinguish even the most divergent categories (‘1’ vs. ‘4+’).\\[-0.25cm]
 
\mybox{\textbf{Finding \#3:} Metrics such as BLUE, ROUGE-L, and METEOR are not a substitute for human judgment, as they still fall short in capturing the nuanced differences inherent in source code comments generation.
Therefore, studies that evaluate results generated by LLMs, especially in the field of software documentation, should not rely solely on the results of such metrics. It is important that these results be complemented by expert judgment.}

\section{Source Code Metrics Analysis}
\label{sec:codeAssessment}

\begin{figure*}[htb]
    \centering
    \begin{tabular}{|c|c|}
        \hline
        \includegraphics[width=0.4\textwidth, trim=3 3 0 0, clip]{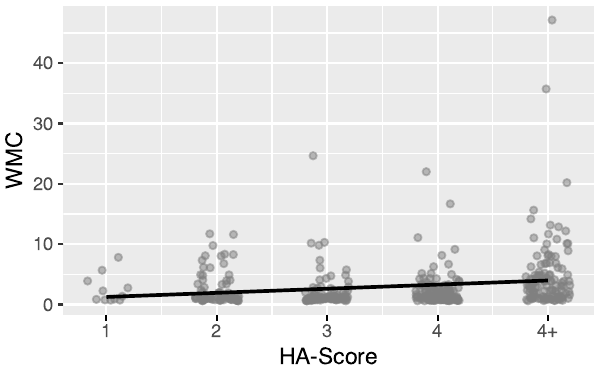} & 
        \includegraphics[width=0.4\textwidth, trim=3 3 0 0, clip]{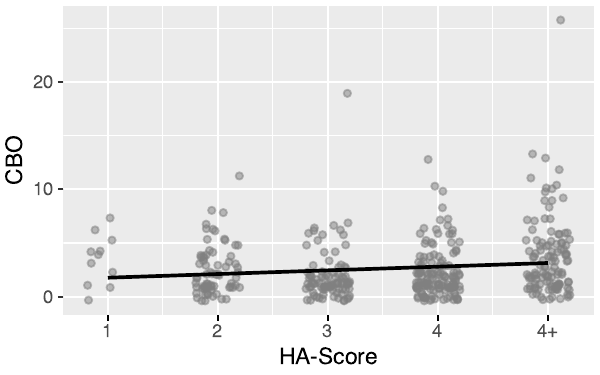} \\ 
        \hline
        \includegraphics[width=0.4\textwidth, trim=3 3 0 0, clip]{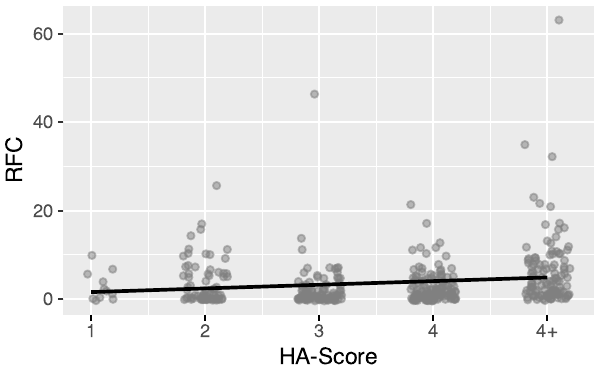} & 
        \includegraphics[width=0.4\textwidth, trim=3 3 0 0, clip]{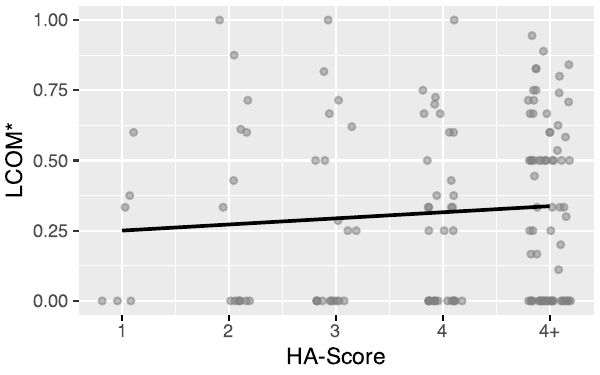} \\
        \hline
        \includegraphics[width=0.4\textwidth, trim=3 3 0 0, clip]{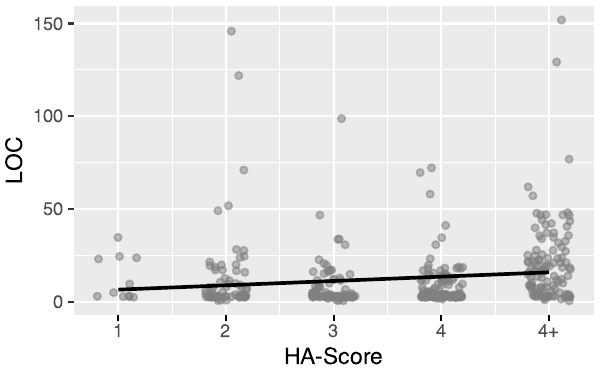} & 
        \includegraphics[width=0.4\textwidth, trim=3 3 0 0, clip]{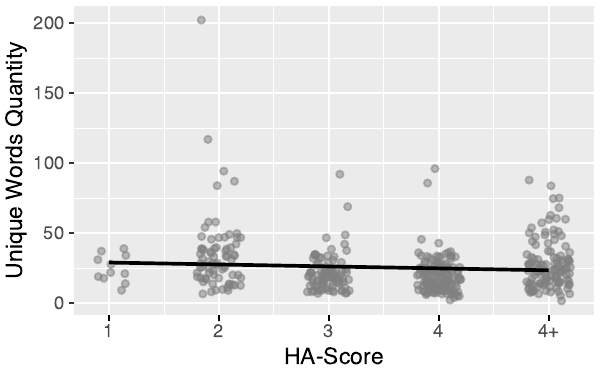} \\ 
        \hline
    \end{tabular}
    \caption{CK-based metrics vs HA-Scores: Spearman correlations with linear regression trends}
    \label{fig:strip_plots}
\end{figure*}

We also investigated which source code properties influence LLMs' ability to produce high-quality source code comments. Figure~\ref{fig:strip_plots} presents the Spearman correlation results between the metrics extracted using the CK Tool~\cite{aniche-ck} and the HA-Score assigned to the corresponding generated Javadoc comments.

Interestingly, some metrics show a slightly positive correlation with the HA-Score. This includes complexity (WMC, Spearman 0.28), size (LOC, Spearman 0.27), and coupling (RFC, Spearman 0.23). Thus, to some extent, number of lines of code, cyclomatic complexity, and richer dependency sets help LLMs generate more accurate comments, likely by providing richer contexts.

Figure~\ref{fig:shortMethodLowHa} illustrates this behavior. The code snippet shows a very short (LOC: 3), simple (WMC: 1), and loosely coupled (RFC: 0) method. Yet, the LLM failed to generate a high-quality Javadoc for it. Compared to the original documentation, the generated version received an HA-Score of 1, the lowest possible rating, indicating that the two are not equivalent but rather opposites. This suggests that providing richer contextual information could have helped the LLM produce higher-quality documentation. Figure~\ref{fig:longerClassHighHa}, on the other hand, shows the opposite case. The class is longer (LOC: 11), moderately complex (WMC: 2), and more tightly coupled (RFC: 8), thus offering the LLM richer structural and contextual cues. As a result, the generated Javadoc not only matched the original but surpassed it in quality, earning a high HA-Score of 4+.

                \begin{figure}[H]
                    \centering
                    \begin{lstlisting}
*bold*Original Javadoc*bold*
/**
 * Configure how unknown String keys will be handled.
 * @param fixedInput if false, an unknown key will be returned from {@link #intValue} as <code>-1</code>; if true,
 * the result will be undefined, and the resulting code will be faster
 */

*bold*LLM-Generated Javadoc*bold*
/**
 * Set whether the input for this method is fixed.
 * @param fixedInput true if the input is fixed, false otherwise
 */

*bold*Code Snippet*bold*
public void setFixedInput(boolean fixedInput) {
    this.fixedInput = fixedInput;
}
                    \end{lstlisting}
                    \caption{Short method (LOC: 3; WMC: 1; RFC: 1) with low-quality LLM-generated Javadoc (HA-Score: 1).}
                    \label{fig:shortMethodLowHa}
                \end{figure}
                
                \begin{figure}[!htbp]
                    \centering
                    \begin{lstlisting}
*bold*Original*bold*
/**
 * {@link Inflector} to for endpoint links.
 */

*bold*LLM-Generated*bold*
/**
 * An inflector that maps container request contexts to responses using an endpoint links resolver.
 */

*bold*CODE*bold*
private static final class EndpointLinksInflector implements Inflector < ContainerRequestContext, Response > {

    private final EndpointLinksResolver linksResolver;

    private EndpointLinksInflector(EndpointLinksResolver linksResolver) {
        this.linksResolver = linksResolver;
    }

    @Override
    public Response apply(ContainerRequestContext request) {
        Map < String, Link > links = this.linksResolver
            .resolveLinks(request.getUriInfo().getAbsolutePath().toString());
        Map < String, Map < String, Link >> entity = OperationResponseBody.of(Collections.singletonMap("_links", links));
        return Response.ok(entity).build();
    }

}
                    \end{lstlisting}
                    \caption{Longer (LOC: 11), moderately complex (WMC: 2), and more tightly coupled (RFC: 8) class with high-quality LLM-generated Javadoc (HA-Score: 4+).}
                    \label{fig:longerClassHighHa}
                \end{figure}

Other metrics exhibit a slightly negative correlation, such as Unique Words Quantity (Spearman -0.05). For example, the mean value of this metric is 35.3 for comments with an HA-Score of 2 and 20.2 for comments with an HA-Score of 4. This suggests that higher lexical diversity does not necessarily aid LLMs in generating better source code comments. Finally, LCOM* shows a minimal and statistically insignificant correlation (Spearman 0.11), indicating that cohesion is not a decisive factor in the quality of the generated comments.\\[-0.25cm]

\mybox{\textbf{Finding \#4:} Size, complexity, and dependencies often indicate poor-quality code. However, we preliminarily find that they tend to provide context that facilitates the automatic generation of comments by LLMs. We note, though, that this positive correlation is not very strong (Spearman $\leq$ 0.30).
}

\section{Threats to Validity}
\label{sec:threatsToValidity}
As with any empirical study, this work faces potential threats to validity~\cite{Experimentation}. First, the HA-Score relies on human judgment, which introduces subjectivity. To mitigate this, both evaluators were experienced software developers with over a decade of combined expertise, and we followed a structured, consistent evaluation process.
Second, the LLMs may have been exposed to the analyzed repositories during training. To reduce this risk, we selected Javadocs only from files created after the model’s training cutoff date. 
Third, the Javadocs used as ground truth may not always reflect high quality comments. To address this threat, both authors independently reviewed the comments, carefully excluding low-quality examples from our ground truth.

\section{Related Work}
\label{sec:relatedWork}

Although they have gained popularity recently, we already have a large number of studies on the use of LLMs in software development activities, including those directly related to source code documentation. For example, \citet{LLMDocumentationAnalysis}~evaluated the use of five LLMs (GPT-3.5, GPT-4, Bard, Llama2, and StarChat) to generate code comments at multiple levels, including inline, function-level, class-level, and folder-level. However, their work differs from ours in two key ways. First, they conducted only a qualitative evaluation, based on criteria such as accuracy, completeness, relevance, understandability, and readability. Second, their study included only 14 code snippets (in Python), whereas we evaluated 415 code snippets (in Java) using both quantitative and qualitative methods.

As another example, \citet{GPT3Documentation}~explored the use of LLMs for automatic comment generation but relied solely on quantitative evaluation using BLEU. The absence of human evaluation or reference comparisons limits the interpretability of their results. In contrast, our study combines multiple quantitative metrics (BLEU, ROUGE-L, METEOR) with structured human assessments, offering a more comprehensive and reliable evaluation framework.

In a recent investigation, \citet{llmEra}~present a large-scale study on LLM-based source code summarization, evaluating four models (GPT-3.5, GPT-4, CodeLlama-Instruct, and StarChat-$\beta$) across ten programming languages. Like our work, they employed BLEU, ROUGE-L, and METEOR for text similarity, and conducted a human evaluation. 
However, their study focuses primarily on function-level code summarization. That is, they do not consider class-level comments, as we did in the present study. Furthermore---and more importantly---their prompts request the generation of summaries limited to a single sentence. For example, the one-shot prompt used in their study is as follows: ``Please generate a short comment {\em in one sentence} for the following function: ⟨code⟩.'' In contrast, we do not limit the generated comments to a single sentence, since our focus is on documentation and not on code summarization. In fact, our prompt explicitly recommends generating comments in the Javadoc format---that is, comments that make use of tags such as {\texttt @param}, {\texttt @see}, etc. Javadoc is the standard format for code documentation in Java and is widely used by developers of this language. Moreover, similar formats exist for other languages, which broadens the applicability of our results to some extent.

\citet{canDevsPrompt}~conducted a controlled experiment to examine how effectively developers can prompt LLMs for documentation generation. They compared ad-hoc versus predefined few-shot prompts and evaluated output quality across six dimensions using human judgment. Their results showed that predefined prompts generally led to more concise and readable documentation, particularly among less experienced users. While their work emphasizes prompt design and user experience, our study complements this by focusing on output quality across a large dataset using both quantitative and qualitative measures.

Finally, \citet{RepoAgent}~proposed a framework for repository-level documentation generation using LLMs, which they evaluated through binary human assessments (i.e., better or not better than human-written comments). Although they found that LLM-generated comments often surpassed manual ones, their evaluation approach lacked granularity. In our study, we address this by using a more nuanced 4-point human rating scale, extended with a fifth category (4+) to better capture subtle differences in quality.

\section{Conclusion, Implications, and Future Work}
\label{sec:conclusionAndFutureWork}

This paper presents the results of a study exploring the use of LLMs to automate source code comment generation. Our evaluation combined quantitative metrics, qualitative human assessments, and source code metrics to assess the quality of LLM-generated Javadocs. The dataset consisted entirely of Javadocs created after the LLMs' training cut-off date, thereby avoiding potential bias from comments already seen during training.

The main implications of our study are as follows.\vspace{0.1cm}

\noindent{\em For practitioners:} Our results show that LLMs can produce high-quality comments, with 58.8\% rated as equivalent to the original ones and 27.7\% rated as superior. Interestingly, source code properties typically associated with poor code---such as larger size and higher complexity---show weak but positive correlations with comment quality, likely because richer context helps LLMs generate more informative and accurate documentation.\vspace{0.1cm}

\noindent{\em For tool builders:} Our results confirm that IDEs can easily provide a simple plug-in for automatic comment generation. Users would only need to select a code fragment and request the generation of a comment by an underlying LLM.\vspace{0.1cm}

\noindent{\em For researchers:} We showed that automated metrics alone are insufficient to fully assess documentation quality. Our results indicate that while metrics such as ROUGE-L and METEOR align better with human judgment than precision-based metrics like BLEU, they still fall short of capturing the nuanced characteristics of high-quality documentation. Therefore, there remains room for developing new metrics that more accurately reflect the quality of LLM-generated source code comments.\\

{\noindent\textbf{Replication Package:} 
We provide a repository containing all prompts, datasets, quantitative and qualitative results, and Python scripts used in this study, along with Docker-based tools and detailed instructions to replicate our experiments or reproduce them with other projects. The package is available at \url{https://doi.org/10.5281/zenodo.17343478}}\\

{\noindent\textbf{Acknowledgments:} This research was supported by grants from CNPq and FAPEMIG.

\bibliographystyle{ACM-Reference-Format}
\bibliography{sample-base}

\end{document}
\endinput